\documentclass[conference,usenatbib]{basi}
\usepackage[british]{babel}
\usepackage{rotating}
\usepackage{dcolumn}
\begin{document}
\title[Garain, Ghosh, Chakrabarti]{Numerical Simulation of Spectral and 
Timing Properties of a Two Component Advective Flow around a Black Hole}
\author[S.~K.~Garain et~al.]%
       {S.~K.~Garain$^1$\thanks{email: \texttt{sudip@bose.res.in}},
       H.~Ghosh$^2$ and S.~K.~Chakrabarti$^{1,2}$\\
       $^1$S N Bose National Centre for Basic Sciences, Block JD, Sector III, Salt Lake, Kolkata 700098, India\\
       $^2$Indian Centre for Space Physics, Chalantika 43, Garia Station Road, Garia, Kolkata 700084, India}

\pubyear{2013}
\volume{8}
\pagerange{11--14}
\date{Received --- ; accepted ---}
\maketitle
\label{firstpage}

\begin{abstract}
We study the spectral and timing properties of a two component advective
flow (TCAF) around a black hole by numerical simulation. Several cases 
have been simulated by varying the Keplerian disk 
rate and the resulting spectra and lightcurves have been produced 
for all the cases. The dependence of the spectral states and quasi-periodic
oscillation (QPO) frequencies on the flow parameters is discussed. We also 
find the earlier explanation of arising of QPOs as the resonance between infall
time scale and cooling time scale remain valid even for Compton cooling.  
\\[6pt]
\end{abstract}
\section{Introduction}\label{s:intro}
The spectral and timing properties of the accretion disk around a 
black hole give away the vital information about understanding the nature 
of the black hole at the centre. 
Assuming the most general accretion flow configuration, namely, two-component 
advective flow (TCAF) model \citep{1995ApJ...455..623C}, we study by numerical
simulation how the spectral and timing properties of an accretion disk around
a black hole are explained. 
\section{Simulation Set Up and Procedure}
\begin{figure}[h!]
\begin{center}
\begin{tabular}{p{6cm}cp{6cm}}
\raisebox{-\height}{\includegraphics[width=5.5cm]{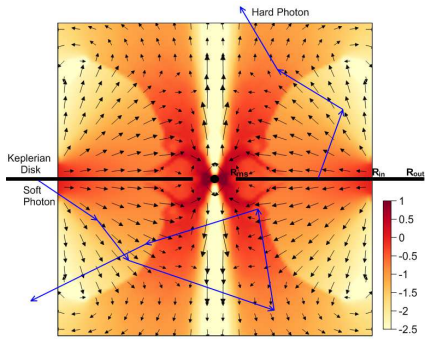}} & \quad &
\caption{\footnotesize{The schematic diagram of our simulation set up. The velocity vectors
of the infalling matter are shown. The colors show the normalized density
in a logarithmic scale. The zig-zag trajectories (blue online) are the
typical path followed by the photons. The velocity vectors are plotted for 
$\lambda=1.73$ and $\epsilon=0.0021$.}
}
\end{tabular}
\end{center}
\end{figure}
\vskip -0.3cm
In Figure 1, the schematic diagram of our simulation set up is presented.
The outer boundary of the sub-Keplerian matter is at $R_{in} = 100 r_g$
($r_g={2GM\over c^2}$) whereas that of the Keplerian disk at the equatorial plane is located
at $R_{out} = 200 r_g$. At the center, a non-rotating black hole of
mass 10$M_{\odot}$ is located. The sub-Keplerian flow dynamics is simulated 
using a TVD code \citep{1996ApJ...470..460M, 2010MNRAS.403..516G}.
For a particular simulation, we use the Keplerian disk rate ($\dot{m}_d$)
and the sub-Keplerian halo rate ($\dot{m}_h$) as parameters. The specific
energy ($\epsilon$) and the specific angular momentum ($\lambda$)
determines the hydrodynamics (shock location, number
density and velocity variations etc.) and the thermal properties
of the sub-Keplerian matter.
The radiative properties of the accretion disk is studied using a Monte
Carlo code \citep{1983ASPRv...2..189P, 2009IJMPD..18.1693G}.
The hydrodynamic code and the radiative transfer code are
coupled together and are run back to back. 
The details of the coupling procedure can be found in \citet{2011MNRAS.416..959G}
and \citet{2012ApJ...758..114G}.
%
\section{Results and Discussions}
\begin{center}
{\footnotesize{
\begin {tabular}[h]{|c|c|c|c|c|c|c|c|}
\hline
\multicolumn{8}{|c|}{Table 1: Parameters used for the simulations and a summary of results.}\\
\hline Case ID & $\epsilon, \lambda$ & $\dot{m}_d$ & $\dot{m}_h$ & ${\rm <R_{sh}>}$ & $\nu_{QPO}$ & $<\alpha>$ & $t_{in}\over t_{cool}$\\
\hline
C1 & 0.0021, 1.73 & 1E-4 & 0.1 & 25.75 & No QPO & 0.826 & 0.694\\
C2 & 0.0021, 1.73 & 2E-4 & 0.1 & 22.82 & 10.63 & 0.853 & 0.844\\
C3 & 0.0021, 1.73 & 3E-4 & 0.1 & 20.39 & 12.34 & 0.868 & 0.954\\
C4 & 0.0021, 1.73 & 4E-4 & 0.1 & 18.62 & 14.63 & 0.873 & 0.944\\
C5 & 0.0021, 1.73 & 5E-4 & 0.1 & 18.33 & 22.74 & 0.901 & 1.071\\
C6 & 0.0021, 1.73 & 1E-3 & 0.1 & 15.02 &  18.2 & 1.074 & 0.993\\
C7 & 0.0021, 1.73 & 1E-2 & 0.1 &  3.4  & No QPO & 1.139 & 3.558\\
C8 & 0.0021, 1.73 & 1E-1 & 0.1 &  3.6  & No QPO  & 1.102 & 33.236\\
\hline
\end{tabular}
}}
\end{center}
In Table 1, we show the parameters used for the simulations and a summary
of the results. In column 5, we present the time averaged shock location 
in $r_g$ near the equatorial plane. For some combinations of $\dot{m}_d$ 
and $\dot{m}_h$, we find QPOs and we list the QPO frequencies (Hz) in column 6. 
The time averaged spectral slope ($\alpha,~I(E)\propto E^{-\alpha}$) is 
given in column 7. In the last column, we show the ratio of the infall 
time scale ($t_{in}$) and the cooling time scale ($t_{cool}$). 

{\bf Spectral properties:}

\begin{figure}
\begin{center}
\begin{tabular}{p{6cm}cp{6cm}}
\raisebox{-\height}{\includegraphics[width=5.5cm]{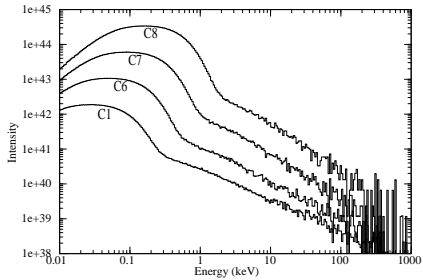}} & \quad &
\caption{\footnotesize{
Variation of the shape of the spectrum when
$\dot{m}_d$ is increased by a factor of 10 starting from $\dot{m}_d = 0.0001$ 
to $0.1$ keeping halo rate constant at $\dot{m}_h=0.1$.
Case IDs are marked for each plot. The spectrum becomes softer as
$\dot{m}_d$ is increased.
}}
\end{tabular}
\end{center}
\end{figure}
\vskip -0.3cm
In Figure 2, we show the variations of the shape of the final emergent
spectra. The case IDs are marked on each plot. As $\dot{m}_d$ is increased, 
the relative intensity increases since increasing $\dot{m}_d$ increases the 
number of soft photons in a given energy band. 
However, the spectra become softer as the centrifugal
pressure dominated inner region (including the post-shock region, when present) 
is cooled faster and the region collapses with the increase of $\dot{m}_d$. 
Also, the number of available hot electrons reduces. 

{ \bf Timing properties:}

\begin{figure}
\begin{center}
\includegraphics[width=5cm,height=5cm]{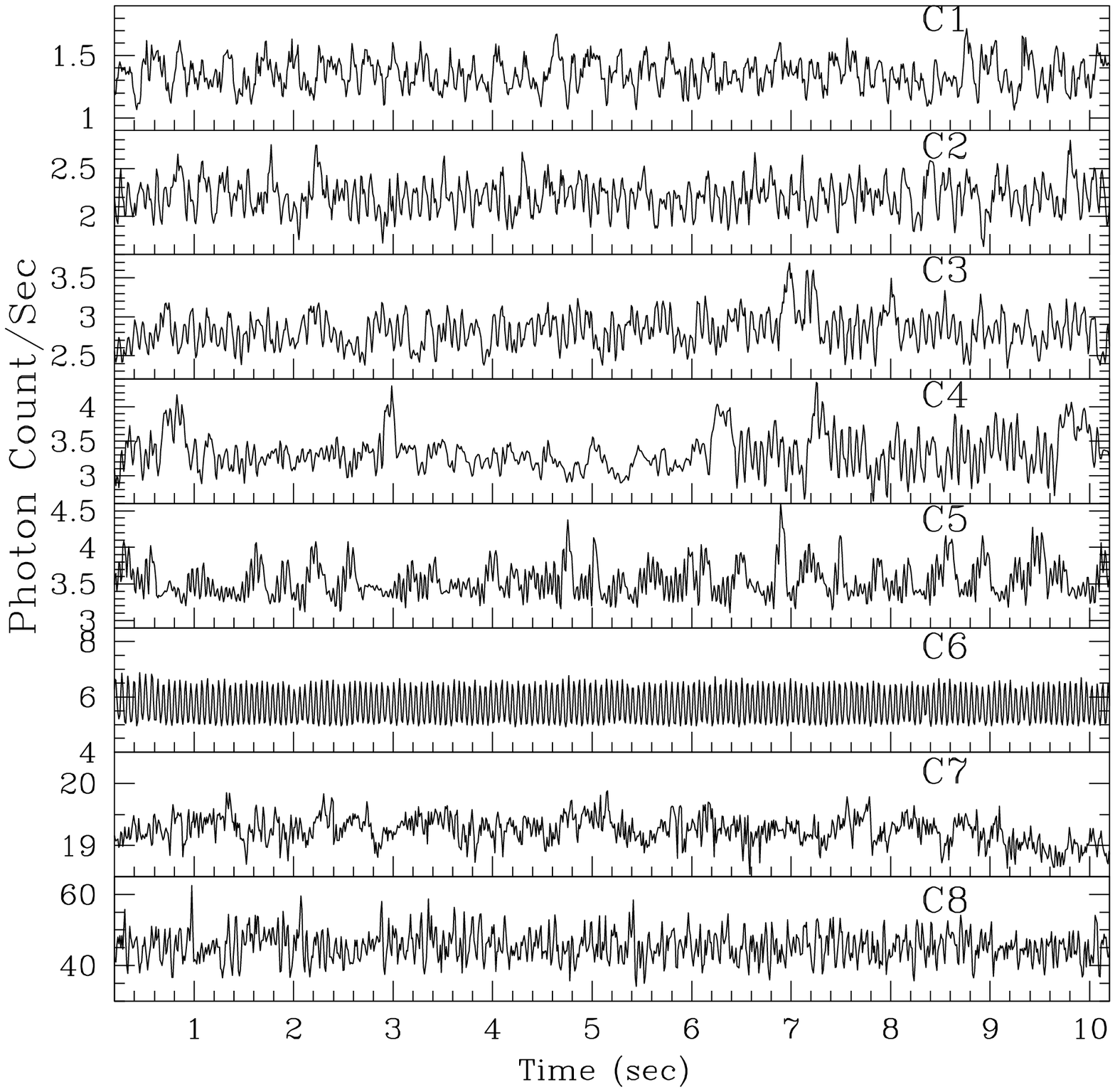}
\includegraphics[width=5cm,height=5cm]{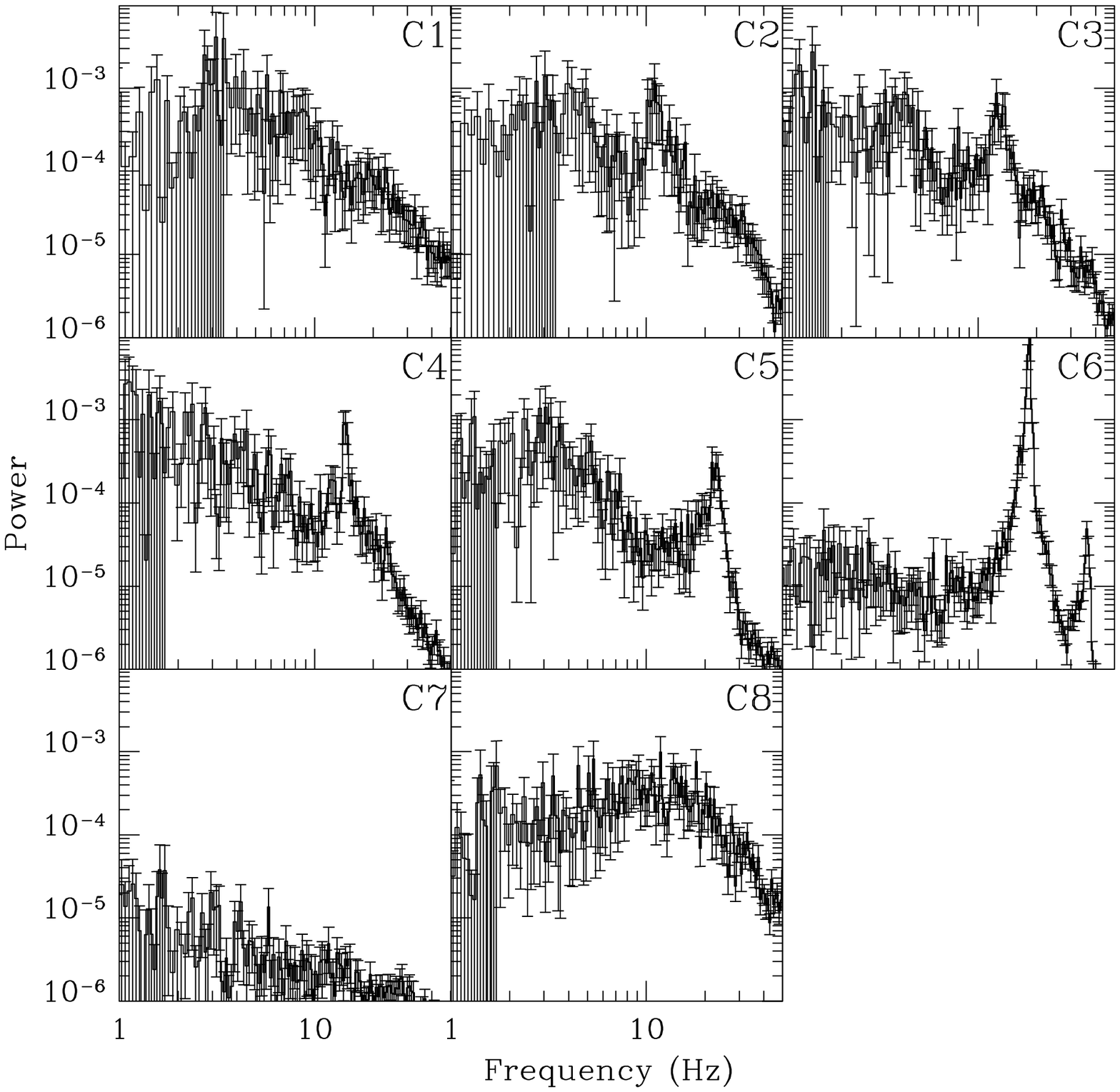}

\hskip 0.5cm (a) \hskip 4.0cm (b)
\caption{\footnotesize {a) The lightcurves of the photons (in the unit of $10^{42}$)
which are in the power-law region ($0.5$keV$<E<100$keV) and b) the
Power Density Spectra (PDS) for these cases are
presented. Here, $\dot{m}_d$ is increased keeping $\dot{m}_h=0.1$ constant.
}}
\end{center}
\end{figure}
\vskip -0.3cm
We compute the time variations of the photon count rates for all the 
cases in order to generate simulated lightcurves.
In Figure 3a, we plot lightcurves of the photons in the energy
band $0.5$keV to $100$keV (for C7 and C8, 2 keV$<E<100$ keV). The photons in
this energy range are mostly the inverse-Comptonized photons. 
The variations in the lightcurves are arising because of
the variations in the hydrodynamic and thermal properties of the post-shock region.
In Figure 3b, we show the Power Density Spectra (PDS) for all the cases.
We find low frequency quasi-periodic oscillations (LFQPO) for some cases. 
The frequencies are listed in Table 1. We find that LFQPO frequencies 
increase with the increase of $\dot{m}_d$.  We have seen that the spectra become
softer with the increase of $\dot{m}_d$. Therefore from Figures 2 and 3b, 
we find that the LFQPO increases as the object transits from the harder 
state to the softer states. 

It has been argued in literature \citep{1996ApJ...457..805M, 2004A&A...421....1C} that the 
oscillation in the centrifugal barrier dominated hot region is responsible for the LFQPO observed in the
black hole candidates. LFQPO arises when the infall time scale of post-shock
matter roughly matches with the cooling time scale. 
For the present case, we compute the infall time scale $t_{in}$ and the
cooling time scale $t_{cool}$, and we present the ratio
$t_{in}\over t_{cool}$ in the last column of Table 1. We see that this ratio is nearly
1 for all the cases when LFQPOs are seen. Thus the proposal of LFQPOs arising out
of resonance oscillation of the post-shock region appears to be justified.
\section*{Acknowledgements}
The work of HG was supported by a post doctoral grant from Ministry of
Earth Science, Government of India. The authors acknowledge the organizers
for providing the local hospitality at Indian Institute of Technology, 
Guwahati, India during the conference.

\label{lastpage}
\end{document}